\documentclass[12pt]{article}
\usepackage{scicite}
\usepackage{times}
\usepackage{amsmath,amssymb}
\usepackage{graphicx}
\usepackage{color}
\usepackage[normalem]{ulem}	
\newenvironment{sciabstract}{%
\begin{quote} \bf}
{\end{quote}}

\newcounter{lastnote}

\title{Programming Curvature using Origami Tessellations}

\author
{Levi H. Dudte$^{1}$, Etienne Vouga$^{1}$, Tomohiro Tachi$^{2}$, L. Mahadevan$^{1,3,4,5\ast}$\\
\\
\footnotesize{$^{1}$Paulson School of Engineering and Applied Sciences, Harvard University}\\
\footnotesize{$^{2}$Graduate School of Arts and Sciences, University of Tokyo, Japan}\\
\footnotesize{$^{3}$Departments of Physics, and Organismic and Evolutionary Biology, Harvard University}\\
\footnotesize{$^4$ Wyss Institute for Bio-inpsired Engineering, Harvard University}\\
\footnotesize{$^5$ Kavli Institute for Nanobio Science and Technology, Harvard University, Cambridge, MA 02138, USA}\\
\footnotesize{$^\ast$To whom correspondence should be addressed; E-mail:  lmahadev@g.harvard.edu.}
}

\date{}

\begin{document} 


\baselineskip24pt


\maketitle 

\begin{sciabstract}
Origami describes rules for creating folded structures from patterns on a flat sheet, but does not prescribe how patterns can be designed to fit target shapes.  Here, starting from the simplest periodic origami pattern that yields one degree-of-freedom collapsible structures - we show that scale-independent elementary geometric constructions and constrained optimization algorithms can be used to determine spatially modulated patterns that yield approximations to given surfaces of constant or varying curvature. Paper models confirm the feasibility of our calculations. We also assess the difficulty of realizing these geometric structures by quantifying the energetic barrier that separates the metastable flat and folded states. Moreover, we characterize the trade-off between the accuracy to which the pattern conforms to the target surface, and the effort associated with creating finer folds. Our approach enables the tailoring of origami patterns to drape complex surfaces independent of absolute scale, and quantify the energetic and material cost of doing so.
\end{sciabstract}

Origami is an art form that likely originated with the invention of paper in China, but was refined in Japan.  The ability to create complex origami structures depends on folding thin sheets along creases, a natural consequence of the large scale separation between the thickness and the size of the sheet. This allows origami patterns to be scaled; the same pattern can be used at an architectural level or at a nanometric level. The richness of the mathematics of origami \cite{Demaine}, together with the promise for technology in the context of creating building blocks for foldable or deployable structures and machines \cite{Lang} has led to an explosion of interest in the subject. Much of the complexity of the folding patterns arise from the combinatorial possibilities associated with the basic origami fold:  the 4-coordinated mountain-valley structure and forms the heart of the simplest origami tessellation depicted in Fig.~\ref{fig:miura-geom}a,c, formed by tiling the plane with a unit cell of four parallelogram tiles and creasing along tile edges. The eponymous Miura-ori was popularized as a structure for solar sail design \cite{miura80}, but the pattern has been used at least since the 15th century, e.g. in the collar pattern of Bronzino's \emph{Portrait of Lucrezia Panciatichi}. It also occurs in many natural settings, including insect wings and leaves \cite{KOBAYASHI98,mahadevan05}, vertebrate guts \cite{shyer13} and is the result of the spontaneous wrinkling of soft adherent thin elastic films \cite{BOWDEN98,RIZZIERI06, AUDOLY08}. Interest in the Miura-ori and allied patterns has recently been rekindled by an interest in mechanical meta-materials ~\cite{wei13,schenk13,waitukaitis15,silverberg14,silverberg15} on scales that range from the architectural to the microscopic.  

\section*{Geometry of  Miura-ori}

The suitability of the Miura-ori for engineering deployable or foldable structures is due to its high degree of symmetry embodied in its periodicity, and four important geometric properties:\emph{(a)} it can be \emph{rigidly folded}, i.e. it can be continuously and isometrically deformed from its flat, planar state to a folded state; \emph{(b)} it has only one isometric degree of freedom, with the shape of the entire structure determined by the folding angle of any single crease; \emph{(c)} it exhibits negative Poisson's ratio: folding the Miura-ori decreases its projected extent in both planar directions; and \emph{(d)} it is \emph{flat-foldable}: when the Miura-ori has been maximally folded along its one degree of freedom, all faces of the pattern are coplanar.

\begin{figure}[h]
	\centering
    \includegraphics[width=\textwidth]{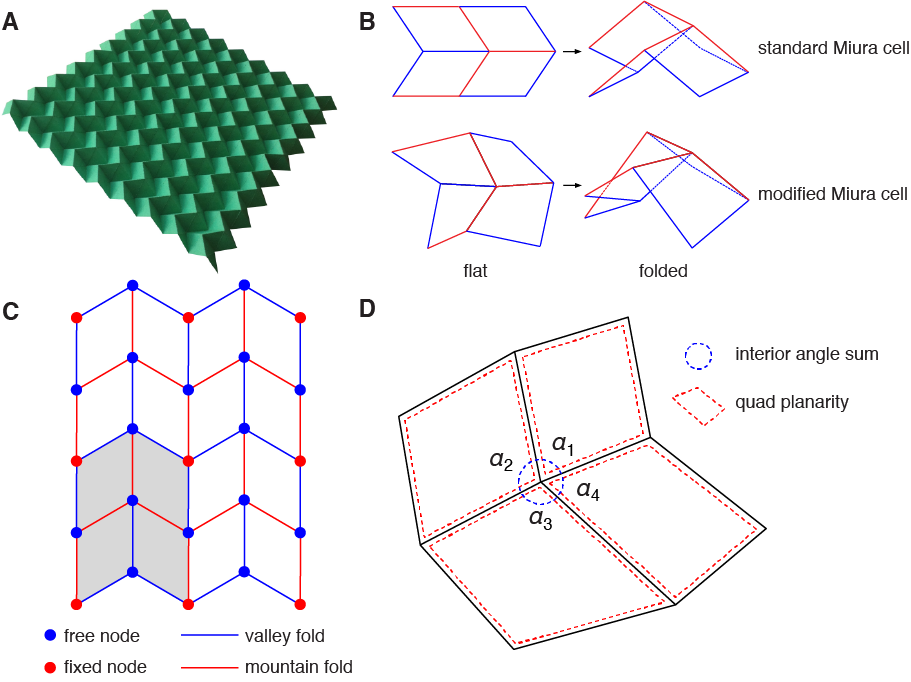}
    \caption{\textbf{Geometry of generalized Miura-ori} (A) Planar periodic Miura-ori  (B) Standard (top) and modified (bottom) Miura-ori unit cells showing the mountain-valley folds  (C) Mountain/valley fold orientations and the pattern of fixed/free nodes for numerical optimization method (D) Constraints at nodes and facets. Facet (quad) planarity implies that the volume of the tetrahedron defined by each quad to vanish. Developability requires that $\sum_i^4 \alpha_i = 2\pi$ and local flat-foldability requires $\alpha_1 + \alpha_3 = \alpha_2 + \alpha_4 = \pi$ (Kawasaki's theorem).}
\label{fig:miura-geom}
\end{figure}

Given the simplicity of the Miura-ori pattern, a natural question is to ask if there are generalizations of it. In particular, for an arbitrary surface with intrinsic curvature, does there exist a Miura-ori-like tessellation of the plane that, when folded, approximates that surface? If so, can this pattern be made rigidly foldable with one degree of freedom? The ability to even partially solve this \emph{inverse problem} would open the door to engineering compact, deployable structures of arbitrary complex geometry, while highlighting the importance of obstructions and constraints that arise when working with materials that {transform} by virtue of their geometric scale separation. { We build on our collective understanding of the geometry of Miura-ori~\cite{tachi09}, mechanics of origami~\cite{wei13,schenk13,silverberg14, silverberg15,waitukaitis15}, existing explorations of the link between fold pattern and geometry~\cite{gattas14} (Fig.~\ref{fig:miura-geom}b), and previous origami \cite{tachi10,tachi13,zhou15} and kirigami\cite{castle14,sussman15,blees15} surface approximations, to pose the inverse problem of fitting Miura-like origami tessellations to surfaces with intrinsic curvature. We then show that the problem can be solved for generalized cylinders using a direct geometric construction and for arbitrarily curved surfaces using a simple numerical algorithm.  Additionally, we characterize the deployability of generic structures, showing how modifications to the geometry of patterns fitting the same target surface effectively tunes their mechanical bistability. Finally, we demonstrate self-similarity of patterns across resolution scales and quantify a trade-off between accuracy and effort involved in surface approximation with origami tessellations.}

Since the periodic Miura-ori pattern tiles the entire plane, we look for generalized origami tessellations, using quadrilateral unit cells that are not necessarily congruent but vary slowly in shape across the tessellation. An embedding of such a pattern in space can be represented as a quadrilateral mesh given by a set of vertices, with edges connecting the vertices and representing the pattern creases, and exactly four faces meeting at each interior vertex. A quadrilateral mesh of regular valence four must satisfy two additional constraints to be an embedding of a generalized Miura-ori tessellation: each face must be planar, and the neighborhood of each vertex must be \emph{developable}, i.e. the interior angles around that vertex must sum to $2\pi$ (Fig.~\ref{fig:miura-geom}d).

\section*{Inverse Origami Design}

A generalized Miura-ori tessellation is guaranteed to possess some, but not all, of the four geometric properties of the regular Miura-ori pattern. An arbitrary unit cell has only one degree of freedom, and this local property guarantees that the global Miura-ori pattern, if it is rigid-foldable at all, must have only one degree of freedom. Moreover, since each unit cell must consist of three valley and one mountain crease, or vice-verse, it must fold with negative Poisson's ratio. Unfortunately, no local condition is known for whether an origami pattern is flat-foldable; indeed it has been shown~\cite{bern96} that the problem of determining global flat-foldability is NP-complete. However, several necessary flat-foldability conditions do exist, of which the two most pertinent are (i) if a generalized Miura-ori tessellation is flat-foldable, each pair of opposite interior angles around each vertex must sum to $\pi$ ~\cite{kawasaki89}  (Fig.~\ref{fig:miura-geom}d), and (ii) if there is a non-trivial generalized Miura-ori embedding (not flat or flat-folded) which satisfies Kawasaki's theorem, it is globally flat-foldable and rigid-foldable~\cite{tachi09}. In practice, enforcing a weaker version of Kawasaki's theorem does improve the degree to which a generalized Miura-ori tessellation is deployable, and in the case where a flat-foldable configuration cannot be found, one can characterize the departure from rigid-foldability by measuring the maximum strain required to deform or snap the bistable tessellation between flat and curved states, a desirable property for stable deployable structures.

These considerations now allow us to formulate the inverse Miura-ori problem: given a smooth surface $M$ in $\mathbb{R}^3$ of bounded normal curvature, an approximation error $\epsilon$, and a length scale  $s$, does there exist a generalized Miura-ori tessellation that \emph{(a)} can be isometrically embedded such that the embedding has Hausdorff distance at most $\epsilon$ to $M$; \emph{(b)} has all edge lengths at least $s$? In particular, do there exist such tessellations that satisfy the additional requirement of being flat-foldable? Less formally, we ask here if it is possible to find optimal Miura-ori tessellations that can be used to conform to surfaces with single or double curvature, i.e. generalized developables, ellipsoids and saddles, and simple pair-wise combinations of these, that might serve as building blocks for more complex sculptures.

{ We illustrate the richness of the solution space by starting with a simple analytic construction for generalized cylinders and a numerical algorithm for generic, intrinsically curved surfaces. The generalized cylinder constructions -- developable surfaces formed by extruding a planar curve along the perpendicular axis --  are guaranteed to be rigid-foldable with 1 DOF and flat-foldable (see SI for details), making them well-suited to applications involving freeform deployable and flat-packed structures (Fig.2a, ~\ref{fig:flat-foldable}a and SI-Movie1).  This is similar to a study published while this work was under review \cite{zhou15}, although the numerical approach therein did not recognize the underlying geometric construction and the ensuing rigid-foldability and flat-foldability of this class of surfaces (see SI)}. For more general surfaces $M$ with intrinsic curvature, we use a numerical optimization algorithm to solve the inverse problem, using the constraints that a quadrilateral mesh approximating $M$ is a generalized Miura-ori if it satisfies a planarity constraint for each face, and a developability constraint at each interior vertex (see Fig.~\ref{fig:miura-geom}d). For a mesh with $V$ vertices and $F \approx V$ faces, there are therefore $3V$ degrees of freedom and only $V+F \approx 2V$ constraints, suggesting that the space of embedded Miura-ori tessellations is very rich; it is therefore plausible that one or more such tessellations that can approximate a given $M$ can be found. Our algorithm allows us to explore this space, constructing tessellations for surfaces of negative, positive, and mixed Gauss curvature. We observe empirically that while surfaces of negative Gauss curvature, such as the helicoid and the hyperbolic paraboloid, readily admit generalized Miura-ori tessellations for a variety of initial guesses for pattern layout, the space of Miura-ori patterns approximating positively-curved surfaces such as the sphere is less rich. Indeed, choosing initial layouts that respect the rotational symmetry of the surface is particularly important for rapid convergence in the latter situation, and also yields surfaces of mixed curvature, such as formed by gluing all pairwise combinations of patches, i.e. $0/+, 0/-, +/-$ curvature, as shown in Fig.~\ref{fig:results}a-f. To realize our results physically, we laser-perforated the patterns on sheets of paper and folded them manually, a process that is currently the rate-limiting step in large scale manufacturability. The results shown in Fig.~\ref{fig:results}g-l, agree well with our calculated shapes. 
 
\section*{Energetic and Material Costs}
{\ In contrast with generalized cylinders, solutions to the numerical optimization problem are only guaranteed to be discrete developable, and are not necessarily flat- or rigid-foldable: the tessellation can be embedded without strain so that it approximates $M$, or so that it is planar, but generally these are isolated states and folding/unfolding the pattern requires snapping through strained configurations (Fig.~\ref{fig:flat-foldable}b,c). To characterize the failure of a generalized Miura-ori tessellation to be rigid-foldable, we use a simple physically based numerical simulation: instead of modeling each quadrilateral face of the pattern as rigid and planar, we divide it into a pair of triangles and modeled as a thin plate with an elastic hinge (see SI for details). Beginning with the folded configuration, we choose one crease in the pattern and incrementally decrease its bending angle from its folded value $\theta=\theta_{\textrm{max}}$ to its flat value $\theta=0$. For each intermediate value of the angle, we allow the pattern to relax to static equilibrium; the strain energy of the equilibrium configuration measures the geometric frustration of that intermediate state (Fig.~\ref{fig:flat-foldable}c). 

To tune this bistability we introduce an inequality constraint in our numerical optimization approach, by replacing Kawasaki's theorem with a tolerance on the residual associated with deviations from flat-foldability given by $|\pi - \alpha_1 - \alpha_3| \le \epsilon_{ff} \ll 1$, as shown in Fig.~\ref{fig:miura-geom}. Since flat-foldability implies rigid-foldability  for non-trivial configurations~\cite{tachi09}, decreasing $\epsilon_{ff}$ is expected to yield Miura-ori patterns that are closer to rigid-foldable. We test this by considering a Miura-ori tessellation approximating a hyperbolic paraboloid. In Fig.~\ref{fig:flat-foldable}c,d, we show that this is indeed the case; {reducing the flat-foldability residual by an order of magnitude} yields a pattern {{that approximates the same target surface,}} but whose energy barrier to folding is half that of the pattern found without the flat-foldability {{restriction}} (see SI-Movie2 for a visualization of the energy barriers as a function of the geometry of folding). To confirm this experimentally, we subjected folded paper hypars with two extreme values of the flat-foldable residuals to a simple tensile test. In Fig.~\ref{fig:flat-foldable}e, we see that the hypar with the larger residual is stiffer, confirming our theoretical predictions (see SI for experimental details).}

Finally, we turn to the accuracy of using folded structures to approximate smooth surfaces. Clearly, as the individual folds become finer the resulting structure will conform more closely to the desired target {{and will require more effort to fabricate}}. To quantify the tradeoff between accuracy and effort, we consider an origami representation of the hyperboloid, shown in Fig.~\ref{fig:trade-off}a, using three different face sizes, each separated by an order of magnitude. A simple cost function associated with the weighted sum of the number of faces and the Hausdorff distance to the smooth surface allows us to follow the minimum cost as a function of the relative weight penalizing effort and accuracy; as expected, when facets are cheap, one can get high accuracy at low cost, but as they become more expensive, for the same cost, accuracy plummets, shown in Fig.~\ref{fig:trade-off}b. Additionally, as the number of facets increases, the area of the folded origami tesselation scaled by the true area of the smooth surface it approximates asymptotically approaches a constant that is larger than that of the actual hyperboloid,  as shown in Fig.~\ref{fig:trade-off}c.

\section*{Outlook}
Our study provides an optimal approximation procedure to solve the inverse problem of determining generalized Miura-ori tessellations that conform to prescribed surfaces. For generalized cylinders, we have shown that the constructed pattern is rigid-foldable and flat-foldable, and thus can be easily adapted to thick origami \cite{chen15}. For doubly-curved surfaces, our computational tool allows us to calculate physically realizable tessellations which we confirm by building paper models. When the Miura-ori tessellations found using our tool are not  flat-foldable, a mechanical model of these surfaces allows us to quantify the strains and energetics associated with snap-through as the pattern moves from the flat to folded configuration. Refining the process allows the folded approximant to approach the smooth target surface which we quantify via a trade-off between accuracy and effort. All together our study opens the way to origamize aribtrary smooth heterogeneously curved surfaces by starting with the simplest origami fold and stitching together an alphabet of generalized Miura-ori tessellations into a flexible design language for engineering shape at any scale.

\section*{Acknowledgments}
We thank the Harvard Microrobotics Lab for help with laser cutting, James Weaver for help with measuring the stress-strain behavior of origami hypars, and the Harvard MRSEC DMR 14-20570, NSF/JSPS EAPSI 2014 (LD), Japan Science and Technology Agency Presto (TT) and the MacArthur Foundation (LM) for partial financial support. 

\section*{Author contributions}
LD, EV and LM conceived and designed the research, with later contributions from TT. LD conducted the simulations and built the models. LD, EV and LM analyzed the results and wrote the manuscript.

\section*{Methods}
\begin{figure}[h]
\centering
    \includegraphics[width=.75\textwidth]{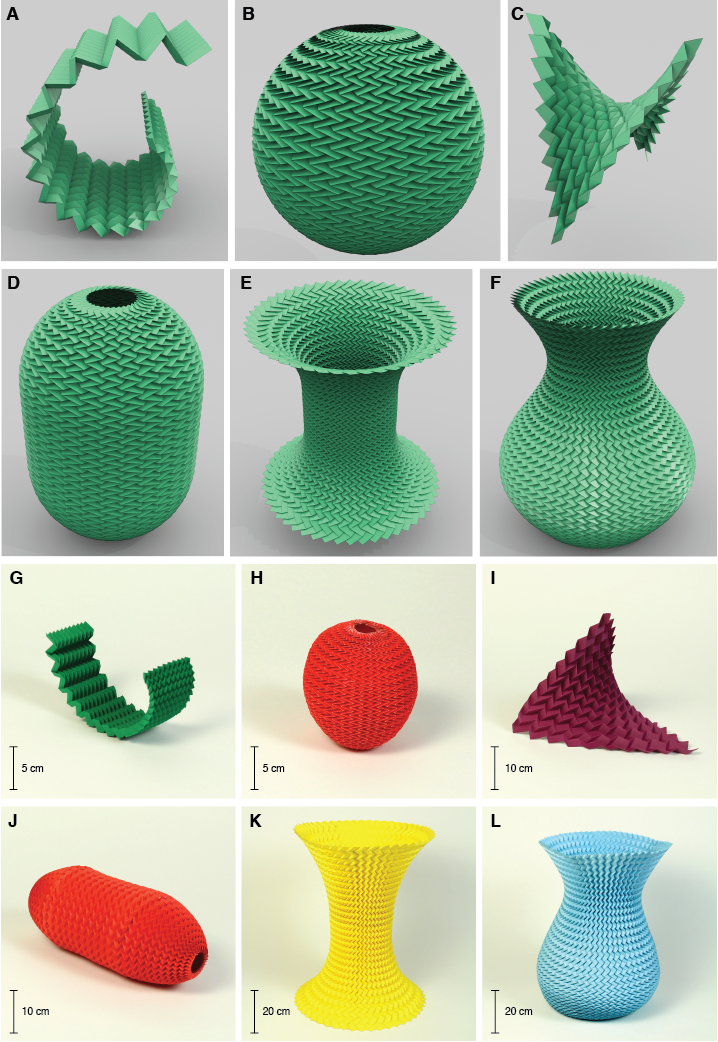}
    \caption{\textbf{Optimal calculated origami tessellations and their physical paper analogs.} (a,g)  Logarithmic spiral - zero Gauss curvature (generalized cylinder) (b,h) Sphere - positive Gauss curvature  (c,i) Hyperbolic paraboloid - negative Gauss curvature (d,j) Pill - cylindrical waist with positively-curved caps (e,k) Candlestick - cylindrical waist with negatively-curved caps (f,l) Vase - positively-curved base with negatively-curved neck.}
\label{fig:results}
\end{figure}

\begin{figure}[h]
	\centering
    \includegraphics[width=0.85\textwidth]{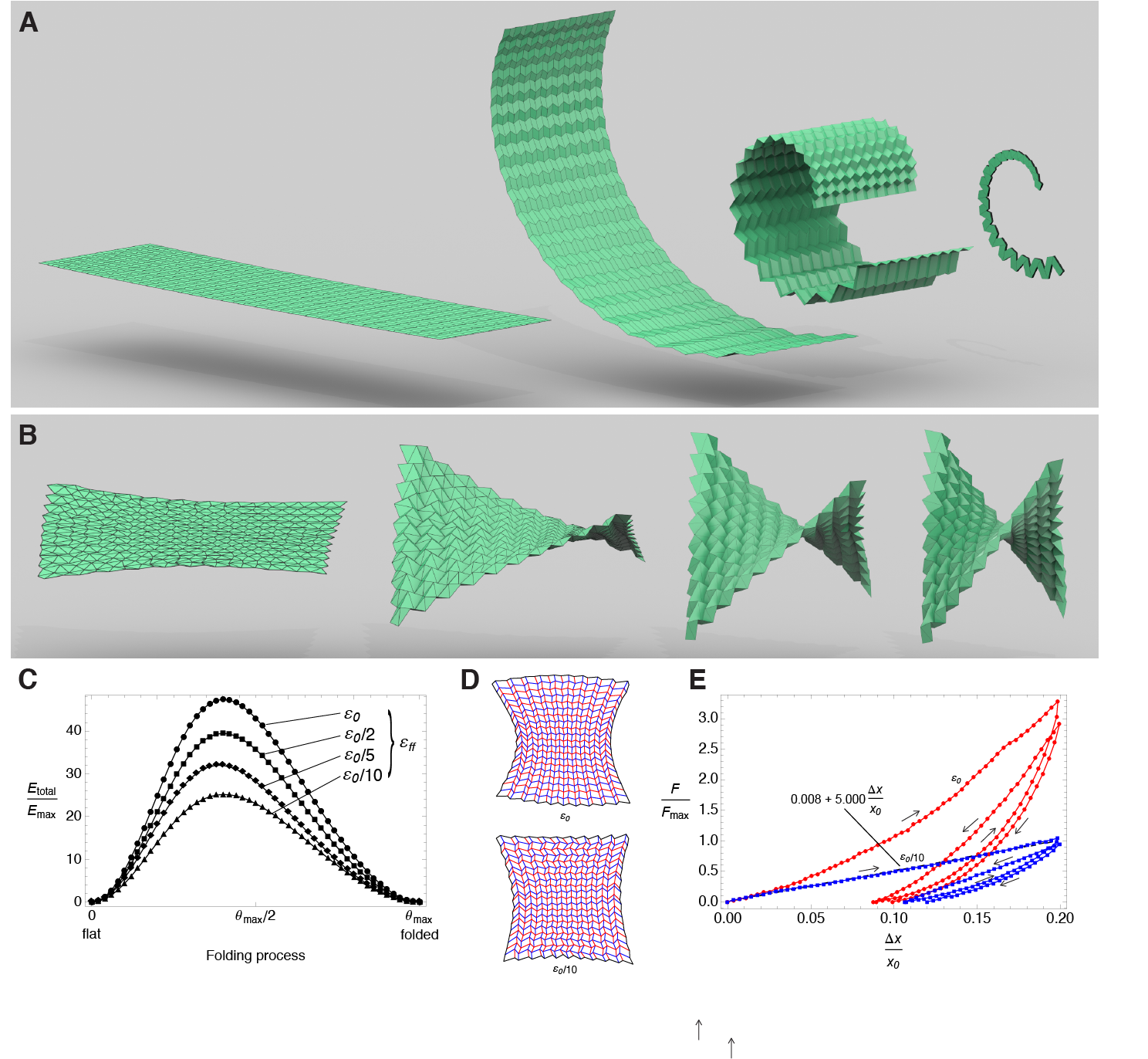}
    \caption{\textbf{Foldability} (a) Generalized cylindrical Miura-ori patterns are rigid-foldable and flat-foldable. The logarithmic spiral here folds rigidly from flat pattern (left) through the target surface and onto the flat-folded plane (right). (b) Generalized Miura-ori patterns solved numerically on doubly-curved surfaces, however, are not rigid-foldable or flat-foldable. We add an extra edge to each planar quad, thereby allowing bending, to deploy these structures. Shown here is the rigid folding of a triangulated hyperbolic paraboloid pattern from flat (left) to solved (right) states. (c) Structures which are not rigid-foldable are bistable with energetic minima at the flat and solved states. For intermediate folding states we minimize the bending of all quads to a non-zero residual strain configuration. Decreasing the flat-foldibility residual $\epsilon_{ff}$ by an order of magnitude effectively halves the magnitude of the energy barrier. (d) Hyperbolic paraboloid patterns with $\epsilon_{ff}=\epsilon_{0}=1.6 \times 10^{-1}$ (left) and $\epsilon_{ff}=\epsilon_{0}/10=1.6 \times 10^{-2}$ (right) where red/blue indicates mountain/valley assignments. The patterns correspond to top and bottom energy curves in (c), respectively. (see SI-Movie2). (e) Force extension experiments on folded paper hypars corresponding to the patterns in (d) confirm that the larger the residual, the higher the stiffness of the resulting structure (red), and thus the higher the barrier separating these bistable structures ($x_0 = 90$mm and $Fmax = 0.431$N). The first experiment with each structure is different owing to the role of some irreversible deformations, but after a couple of cycles, the force-extension characteristic settles onto a reproducible curve.}
\label{fig:flat-foldable}
\end{figure}

\begin{figure}[h]
	\centering
    \includegraphics[width=0.95\textwidth]{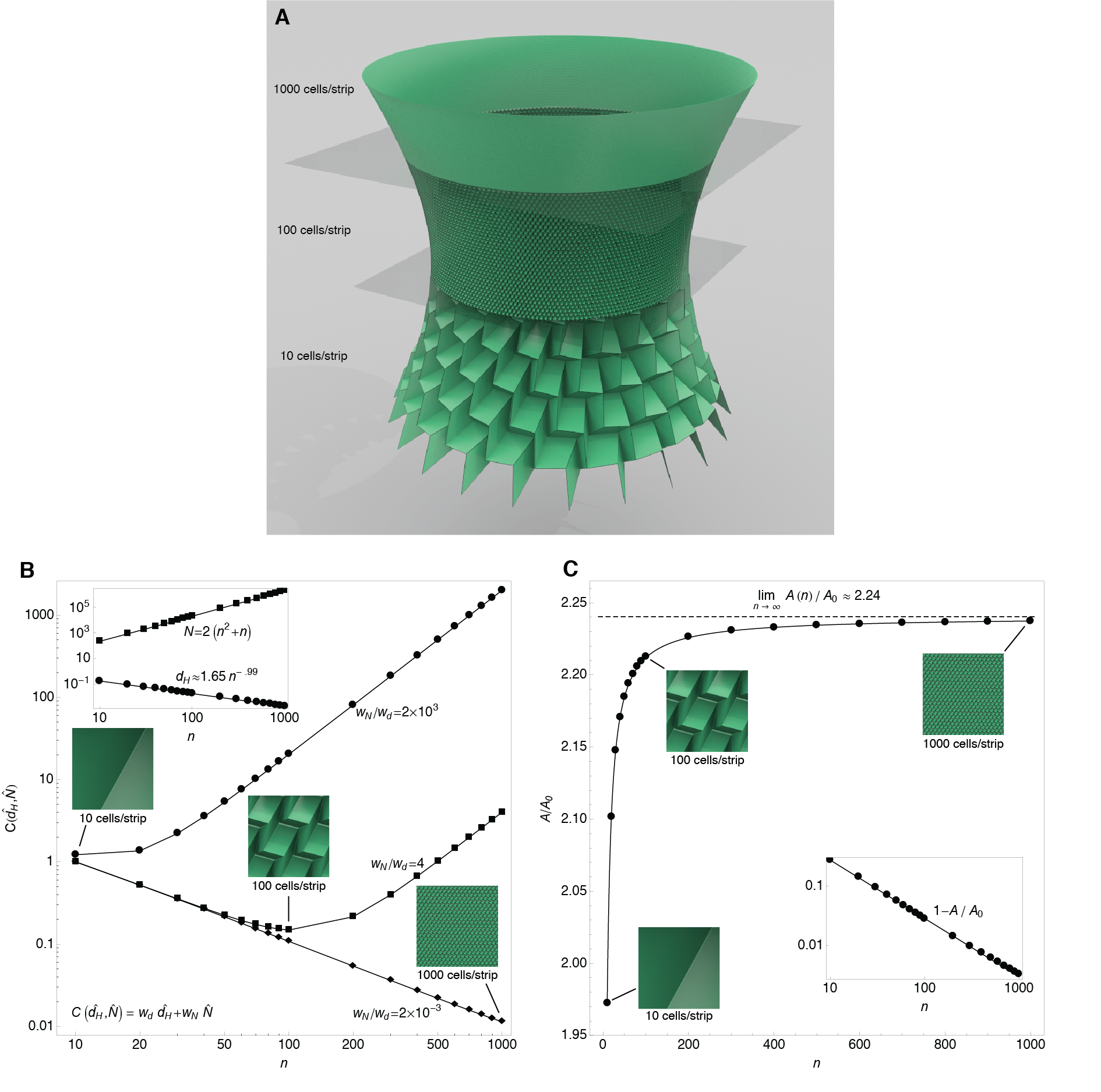}
    \caption{\textbf{Accuracy-effort tradeoff in origami tessellations} (a) Three Miura-ori approximations of a hyperboloid, shown in part, differ from each other by a factor of 10. Increasing the density of facets allows us to approach the smooth hyperboloid. (b) A simple cost function that  is the sum of the number of faces and the Hausdorff distance to the smooth hyperboloid shows a clear minimum: as the cost of facets (independent of their area) increases, the optimum shifts towards the coarse approximation and as the facets become cheaper, the optimum shifts towards the finer approximation. (c) The non-dimensional area of the curved Miura-ori approximation to the hyperboloid (normalized by the area of the smooth hyperboloid) $A/ A_0$ as the number of facets increases approaches a constant greater than unity. In the inset, we see that $(A_0-A) \sim n^{-1}$ consistent with the fact that the facets are self-similar.}
\label{fig:trade-off}
\end{figure}

{\bf Experiment.} We fabricated paper models by perforating patterns with a laser-cutter and folding by hand. Despite originating from a single contiguous pattern, larger models (Fig.~\ref{fig:results}k,l) had to be divided into patches before begin perforated, folded and glued together to form the final surface. The stiffness of the hypar was determined by subjecting the model to a simple force extension test, cycling through increasing the strain of connection points separated by 90 mm to 0.2 and then decreasing the strain until the model reached a new plastic configuration with 0 N incident force. See supplementary information for further details.
\linebreak
{\bf Numerical computations.} The numerical computations (fitting generic surfaces and analyzing energetics) were conducted with custom Matlab code. Analytic gradients of the constraints and objective functions are provided to the \texttt{fmincon} constrained optimization routine and constraint residuals are minimized to at most $1e{-10}$. Surfaces of revolution permitted periodic developability constraints to be applied to a symmetric strip. See supplementary information for further details on constraint patterns, initial conditions and convergence.
\linebreak
{\bf Code availability.} The surface-fitting code is currently patent-pending, details of which are available in the supplementary information. Patterns presented in Fig.~\ref{fig:results} available upon request.

\clearpage

\clearpage

\begin{thebibliography}{10}

\bibitem{Demaine} E. Demaine and J. O'Rourke, ``Geometric folding algorithms: linkages, origami, polyhdera," Cambridge University Press, 2011.

\bibitem{Lang} R.~ Lang, ``Origami design secrets," 2nd edn., CRC Press, 2011.

\bibitem{miura80}K.~Miura, ``Method of packaging and deployment of large membrane in space,'' {\it {Proceedings of the 31st Congress of the International Astronautical Federation}}, New York, pp. 1-10 (1980).


\bibitem{KOBAYASHI98}H.~Kobayashi,~B.~Kresling,~J.~Vincent, ``The geometry of unfolding tree leaves,'' {\it{Proceeding of the Royal Society B: Biological Sciences}} {\bf{265}} {1391, pp. 147-154 (1998).}

\bibitem{mahadevan05}L.~Mahadevan,~S.~Rica, ``Self-organized origami,'' {\it{Science}} {\bf{307}} pp. 1740 (2005).

\bibitem{shyer13} A.~Shyer {\it{ et al.}} "Villification: how the gut gets its villi, " {\it{Science}} {\bf{342}} pp. 212-18 (2013).

\bibitem{BOWDEN98}N.~Bowden,~S.~Brittain,~A.~G.~Evans,~J.~Hutchinson,~G.~Whitesides, ``Spontaneous formation of ordered structures in thin films of metals supported on an elastomeric polymer,'' {\it{Nature}} {\bf{393}}, pp. 146-149 (1998).

\bibitem{RIZZIERI06}R.~Rizzieri,~L.~Mahadevan,~A.~Vaziri,~A.~Donald, ``Superficial wrinkles in stretched, drying gelatin films,'' {\it{Langmuir}} {\bf{22}} {8, pp. 3622-3626 (2006).}

\bibitem{AUDOLY08}B.~Audoly,~A.~Boudaoud, ``Buckling of a stiff film bound to a compliant substrate -- Part III: Herringbone solutions at large buckling parameter,'' {\it{Journal of the Mechanics of Physics and Solids}} {\bf{56}}, pp. 2444-2458 (2008).

  
\bibitem{wei13}Z.~Y.~Wei,~Z.~V.~Guo,~L.~Dudte,~H.~Y.~Liang,~L. Mahadevan, ``Geometric mechanics of periodic pleated origami,'' {\it{Physical Review Letters}} {\bf{110}} {215501 (2013).}

\bibitem{schenk13}M.~Schenk,~S.~D.~Guest, ``Geometry of Miura-folded metamaterials,'' {\it{Proceedings of the National Academy of Sciences}} {\bf{110}} {9 (2013).}

\bibitem{silverberg14}J.~L.~Silverberg, {\it{et. al.}}, ``Using origami design principles to fold reprogrammable mechanical metamaterials,'' {\it{Science}} {\bf{345}} 6197, pp. 647-650 (2014).

\bibitem{silverberg15}J.~L.~Silverberg, {\it{et. al.}}, ``Origami structures with a critical transition to bistability arising from hidden degrees of freedom,'' {\it{Nature Materials}} {\bf{14}}, pp. 389-393 (2015).

\bibitem{waitukaitis15}S.~Waitukaitis,~R.~Menaut,~B.~Chen,~M. van Hecke, ``Origami multistability: From single vertices to metasheets,'' {\it{Physical Review Letters}} {\bf{114}} {055503 (2015).}

 
 \bibitem{tachi09}T.~Tachi, ``Generalization of rigid foldable quadrilateral mesh origami,'' {\it{Proceedings of the International Association for Shell and Spatial Structures (IASS) Symposium}} (2009).

\bibitem{gattas14}M.~Gattas,~Z.~You, ``Miura-based rigid origami: parametrizations of curved-crease geometries,'' {\it{Journal of Mechanical Design}} {\bf{136}} {12 (2014).}

\bibitem{tachi10}T.~Tachi, ``Origamizing polyhedral surfaces,'' {\it{IEEE Transactions on Visualization and Computer Graphics}} {\bf{16}} {2, pp. 298-311 (2010).}

\bibitem{tachi13}T.~Tachi, ``Freeform origami tessellations by generalizing Resch's patterns,'' {\it{Journal of Mechanical Design}} {\bf{135}} {11 (2013).}

\bibitem{zhou15}X.~Zhou,~H.~Wang,~Z.~You, ``Design of three-dimensional origami structures based on vertex approach,'' {\it Proceedings of the Royal Society A} {\bf 471} 2181 (2015).

\bibitem{castle14}T.~Castle, {\it{et. al.}}, ``Making the cut: lattice kirigami rules,'' {\it{Physical Review Letters}} {\bf{113}} {245502 (2014).}

\bibitem{sussman15}D.~Sussman, {\it{et. al.}}, ``Algorithmic lattice kirigami: a route to pluripotent materials,'' {\it{Proceedings of the National Academy of Sciences}} {\bf{112}} {24, pp. 7449-7453 (2013).}

\bibitem{blees15}M.~K.~Blees, {\it{et. al.}}, ``Graphene kirigami,'' {\it{Nature}} (2015).

 
\bibitem{bern96}M.~W.~Bern,~B.~Hayes, ``The complexity of flat origami,'' {\it{Proceedings of the Seventh Annual (ACM-SIAM) Symposium on Discrete Algorithms}}, Atlanta, GA, pp. 175-183 (1996).
 
\bibitem{kawasaki89}T.~Kawasaki, ``On the relation between mountain-creases and valley-creases of a flat origami,'' {\it{Proceedings of the 1st International Meeting on Origami Science and Technology}} (Ed. H. Huzita), Ferrara, Italy, pp. 229-237 (1989).

\bibitem{chen15}Y.~Chen,~R.~Peng,~Z.~You, ``Origami of thick panels,'' {\it{Science}} {\bf{349}} 6246, pp. 396-400 (2015).

\end{thebibliography}
\end{document}